\documentclass[runningheads]{llncs}

\usepackage{amsmath}
\usepackage{graphicx}
\usepackage{xcolor}
\usepackage{multirow}
\usepackage{float}
\usepackage{caption}
\usepackage{subcaption}
\usepackage{hyperref}
\usepackage{booktabs}
\usepackage{enumitem}

\graphicspath{{imgs}}

\usepackage{array}
\newcolumntype{L}[1]{>{\raggedright\let\newline\\\arraybackslash\hspace{0pt}}m{#1}}
\newcolumntype{C}[1]{>{\centering\let\newline\\\arraybackslash\hspace{0pt}}m{#1}}
\newcolumntype{R}[1]{>{\raggedleft\let\newline\\\arraybackslash\hspace{0pt}}m{#1}}

\begin{document}

\title{Deep Learning for Brain Tumor Segmentation in Radiosurgery: Prospective Clinical Evaluation}
% \titlerunning{title running}
\titlerunning{Deep Learning for Radiation Therapy: Clinical Evaluation}

% \author{
%     Anonymous Author \inst{1}
% }
\author{
    Boris Shirokikh \inst{1, 2, 3} \and
    Alexandra Dalechina \inst{4} \and
    Alexey Shevtsov \inst{2, 3} \and
    Egor Krivov \inst{2, 3} \and
    Valery Kostjuchenko \inst{4} \and
    Amayak Durgaryan \inst{5} \and
    Mikhail Galkin \inst{5} \and
    Ivan Osinov \inst{4} \and
    Andrey Golanov \inst{5} \and
    Mikhail Belyaev \inst{1, 2}
}

% \authorrunning{Anonymous Author et al}
\authorrunning{B.Shirokikh et al}

% \institute{
%     Anonymous Institution
%     \\
%     \email{anonymous@email}
% }
\institute{
    Skolkovo Institute of Science and Technology, Moscow, Russia
    \and
    Kharkevich Institute for Information Transmission Problems, Moscow, Russia
    \and
    Moscow Institute of Physics and Technology, Moscow, Russia
    \and 
    Moscow Gamma-Knife Center, Moscow, Russia
    \and 
    Burdenko Neurosurgery Institute, Moscow, Russia
    \\
    \email{m.belyaev@skoltech.ru}
}
\maketitle              % typeset the header of the contribution
\begin{abstract}
Stereotactic radiosurgery is a minimally-invasive treatment option for a large number of patients with intracranial tumors. As part of the therapy treatment, accurate delineation of brain tumors is of great importance. However, slice-by-slice manual segmentation on T1c MRI could be time-consuming (especially for multiple metastases) and subjective. In our work, we compared several deep convolutional networks architectures and training procedures and evaluated the best model in a radiation therapy department for three types of brain tumors:  meningiomas, schwannomas and multiple brain metastases. The developed semiautomatic segmentation system accelerates the contouring process by $2.2$ times on average and increases inter-rater agreement from $92.0$\% to $96.5\%$.

\end{abstract}
\section{Introduction}

% Stereotactic radiosurgery is one of the most non-invasive and efficient ways to treat multiple brain metastases \cite{yamamoto2014stereotactic}. The treatment planning procedure consists of two steps: delineation of all clinically relevant lesions on an MRI scan and the subsequent designing of a treatment plan in order to deliver a prescribed dose of radiation to all targets. Delineation is a particularly labor-intensive process and requires a lot of time and concentration.

Brain stereotactic radiosurgery involves an accurate delivery of radiation to the delineated tumor. The basis of the corresponding planning process is to achieve the maximum conformity of the treatment plan. Hence, the outcome of the treatment is highly dependent on the clinician's delineation of the target on the MRI. Several papers have been shown that experts defined different tumour volumes for the same clinical case \cite{roques2014patient}. As there are no margins applied to a contoured target, the differences in contouring could increase normal tissue toxicity or the risk of recurrence.

The process of contouring is the largest source of potential errors and inter-observer variations in target delineation \cite{torrens2014standardization}. Such variability could create challenges for evaluating treatment outcomes and assessment of the dosimetric impact on the target. Routinely the targets are delineated through slice-by-slice manual segmentation on MRI, and an expert could spend up to one hour delineating an image. However, stereotactic radiosurgery is one-day treatment and it is critical to provide fast segmentation in order to avoid treatment delays. 

Automatic segmentation is a promising tool in time savings and reducing inter-observer variability of target contouring \cite{sharp2014vision}. Recently deep learning methods have become popular for a wide range of medical image segmentation tasks.
In particular, gliomas auto-segmentation methods are well-developed \cite{bratsmeta} thanks to BRATS datasets and contests \cite{BRATS}. At the same time, the most common types of brain tumors treated by radiosurgery, namely meningiomas, schwannomas and multiple brain metastases, are less studied. Recently published studies \cite{charron2018automatic,liu2017deep,krivov2018tumor} developed deep learning methods for automatic segmentation of these types of tumors. However, these studies do not investigate the above-mentioned clinical performance metrics: inter-rater variability and time savings.

Our work aimed to fill this gap and evaluate the performance of semi-automatic segmentation of brain tumors in clinical practice. We developed an algorithm based on deep convolutional neural network (CNN) with suggested adjustment to cross-entropy loss, which allowed us to significantly boost quality of small tumors segmentation. The model achieving the state-of-the-art level of segmentation was integrated into radiosurgery planning workflow. Finally, we evaluated the quality of the automatically generated contours and reported the time reduction using these contours within the treatment planning.

\section{Related work}
\label{sec:related_work}

During recent years, various deep learning architectures were developed.
% U-Net \cite{unet}, one of the most successful recent CNN, was designed for 2D image segmentation.
For medical imaging, the best results were achieved by 3D convolutional networks: 3D U-Net \cite{unet3d} and V-Net \cite{milletari2016v}. However, a large size of brain MRI for some tasks places additional restrictions on CNN. A network called DeepMedic \cite{kamnitsas2017efficient} demonstrated solid performance in such problems, including glioma segmentation \cite{bratsmeta}. % (acute ischemic stroke segmentation \cite{ISLES} and)

Some image processing methods were proposed for the other brain tumors as well. For example, authors of \cite{liu2016automatic} developed a multistep approach utilizing classical computer vision tools such as thresholding or super-pixel clustering.
%Another popular strategy relies on template matching, e.g., \cite{ambrosini2010computer}.
In common with other medical image processing tasks, such methods have two key drawbacks: processing speed and quality of small lesions segmentation \cite{liu2017deep}. Deep learning-based approaches may potentially resolve these issues thanks to its high inference speed and great flexibility. Indeed, several recently published studies validated CNN in the task of nonglial brain tumors segmentation and demonstrated promising results. In \cite{liu2017deep} authors modified the DeepMedic to improve segmentation quality. Authors of \cite{charron2018automatic} compared various combinations of T1c, T2 and Flair modalities. New patch generation methods were proposed and evaluated on three types of brain tumors in \cite{krivov2018tumor}. In \cite{milletari2016v} authors introduced a novel loss function based on Dice coefficient to improve segmentation results in highly class imbalance tasks.

\section{Data}
\label{sec:data}

% \begin{figure}
%     \begin{center}
%       \includegraphics[width=0.95\linewidth]{imgs/hist_mod_diam.png}
%       \caption{}
%       \label{fig:diameters}
%     \end{center}
% \end{figure}

For computational experiments, we used 548 contrast-enhanced T1-weighted MRI with $0.94 \times 0.94 \times 1$ mm image resolution. These cases were characterized by multiple brain tumors ($4.5$ per patient) of different sizes: from $1.3$ mm up to $4.2$ cm in diameter. These images were naturally divided into two datasets. The first one, \emph{training} dataset, consisted of 489 unique patients examined before 2017. It was used to train different models and tune their parameters via cross-validation. The second, \emph{hold-out} dataset, was represented by another 59 patients who were treated in 2017. We performed the final comparison of the best methods on the hold-out dataset to avoid overfitting.

Finally, to evaluate the quality of tumor delineation algorithm in clinical practice, we used the third, \emph{clinical}, dataset which consists of four cases of meningioma, two cases of vestibular schwannoma and four cases of multiple brain metastases (ranged from 3 to 19 lesions per case) collected in 2018. Four experts (or users) with experience in brain radiosurgery ranged from 3 to 15 years delineated each of these cases in two setups: manually and using the output of our model as the starting point, see the details in \ref{sec:clinical_exp}.

% OLD VERSION: To evaluate the quality of tumor delineation algorithm in clinical practice, we obtained data with 10 patients from current workflow of Gamma Knife center. Four experts (or users) delineated these cases with and without usage of our algorithm for the total of 28 tumors, see details in \ref{sec:clinical_exp}.

% for the total of 34 (stats were obtained only on 28 of them) tumors. 

% we asked four experts from Gamma Knife center to use it on their patients. Each expert had to draw contours for each patient. First, without using our delineation algorithm, then the experts were given the CNN-initialized contours and they were able to adjacent the results if they want. So that, we obtained another dataset, that consisted 10 unique patients with total \todo{?} lesions.

\section{Methods}

\subsection{CNN}
\label{sec:network}
We used vanilla 3D U-Net, V-Net and DeepMedic models as network architectures. We trained all models for $100$ epochs, starting with learning rate of $0.1$, and reducing it to $0.01$ at the epoch $90$. Each epoch consists of 200 stochastic gradient descent iterations. At every iteration, we generated training patches of size $64 \times 64 \times 64$ with batches of size $12$ for 3D U-Net and $16$ for V-Net. For DeepMedic we generated $16$ patches of effective size $39 \times 39 \times 39$ in one batch. We used 5-fold cross-validation to split our training data patient-wise. After the train-predict process, we gather test predictions over the 5 splits to form the metric curve and compare experiment results.

For a subset of experiments (see Sec. \ref{sec:results} for the details), we also used a modified loss function, described in the next subsection and Tumor Sampling from \cite{krivov2018tumor}. For the Tumor Sampling as well as the original patches sampling procedures we set the probability to choose the central voxel of each patch belonging to the target mask to be $0.5$ for all experiments. We reported the results on the hold-out dataset while using the training dataset to fit the models.

% For the tumor segmentation task we use a 3D U-Net architecture with several minor changes. We replaced every double convolution by residual blocks \cite{resnet}. We used channel-wise summation instead of concatenation and applied convolutions with kernel $1 \times 1 \times 1$ to the shortcuts before this summation. The number of filters was reduced to $16$ in the initial resolution. We compared our model with vanilla 3D U-Net and V-Net and also vanilla DeepMedic via cross-validation on the training dataset (fig. \ref{fig:models}).

% \begin{figure}[h!]
% \centering
%   \begin{subfigure}[t]{.5\linewidth}
%     \centering
%     \includegraphics[width=\textwidth]{imgs/models.png}
%     \caption{CNN models comparison}\label{fig:models}
%   \end{subfigure}%   
%   \begin{subfigure}[t]{.5\linewidth}
%     \centering
%     \includegraphics[width=\textwidth]{imgs/unet_ts_iwbce.png}
%     \caption{The best model with Tumor Sampling and then with iwBCE }\label{fig:iwbce}
%   \end{subfigure}

%   \caption{\todo{PRC for different experiments}}\label{fig:prc}
% \end{figure}

% \subsection{Motivation}
% All the approaches missed lots of small tumors or inappropriate segmented them. We assumed that such a performance comes from loss function properties: errors on small targets impact to loss function equal to small inaccuracies of large target predictions. To make all possible errors contribute equally to the loss function, we construct a tensor of weights, which are equal to inverse relative volumes of regions of interest.

\subsection{Inversely weighted cross-entropy}
\label{sec:iwbce}

We observed that all methods were missing lots of small tumors or inappropriate segmented them. We assumed that such a performance comes from loss function properties: errors on small targets have the same impact on the loss function as small inaccuracies in large lesions. To make all possible errors contribute equally to the BCE (binary cross-entropy) loss function, we construct a tensor of weights, which are equal to inverse relative volumes of regions of interest.

Given the ground truth on the training stage, we generate a tensor of weights for every image in the train set. To form such a tensor for the given image we split the corresponding ground-truth mask into connected components $C_i, i\in \{0..K\}$, where $C_0$ is the background and $K$ is the number of tumors. Weights of the background component were set to be $w_0 = 1$. The weights for pixels in the connected component $C_i$ $\left( i \neq 0 \right)$ are equal to:

\begin{equation}
    w_i = \beta \cdot \frac{\sum_{k=0}^K |C_k|}{|C_i|},
    \label{eq:weight}
\end{equation}
where $\beta$ is the fraction of positive class in the current training set. %The approximate order of $\beta$ was $10^{-3}$. 
The final form of our loss is the same with weighted BCE over $n$ voxels in the propagated sample:

\begin{equation}
    \text{iwBCE} = - \frac{1}{n} \sum_{j=1}^n \omega_j \cdot \left( y_j \log p_j + \left( 1 - y_j \right) \log \left( 1 - p_j \right) \right),
\end{equation}

where $\omega_j$ is the weight of the $j$-th pixel calculated using (\ref{eq:weight}).

We compare proposed loss function with the current state-of-the-art Dice loss \cite{milletari2016v} as well as with the standard BCE.

\subsection{Metric}
\label{sec:metrics}

We highlighted two essential characteristics that could characterize small tumors segmentation: tumor delineation and detection quality. Since delineation could be simply measured by local Dice score and experts could always adjust contours of found tumors, we focus our attention on the detection quality. 

We suggested measuring it in terms of tumor-wise precision-recall curves. We adopted the FROC curve from \cite{van2010comparing} by changing its hit condition between predicted and ground truth tumors. Predicted tumors were defined as connected components above the probability of $0.5$, and we treated the maximum probability of a component as a model's certainty level for it. Our hit condition is that the Dice score between real and predicted lesions is greater than zero. We found such lesion-wise PRC (precision-recall curve) to be more interpretable and useful for model comparison than traditional pixel-wise PRC.

%We can plot tumor-wise precision-dice curves, capturing both detection and delineation quality. To do that, we perform the procedure defined previously, but now for each threshold value we compute average dice score of all real tumors. Tumor's dice score is considered to be $0$, if there was no prediction above current threshold level for this lesion. Hence, such a plot show that for each precision level, we can provide prediction with observed average dice score. Such plots capture trade-off between tumor-wise precision and average dice score.

% \section{Setup of Experiments}

% \subsection{Neural networks training}
% \label{sec:training}

% We trained our models for $100$ epochs, starting with learning rate of $0.1$, and reducing it to $0.01$ at the epoch $90$. Each epoch consists of 200 stochastic gradient descent iterations. At every iteration we generated training patches of size $64^3$ with batch of size $12$ for 3D U-Net, $16$ for V-Net and $32$ for our U-Net model. For DeepMedic we generated $16$ patches of effective size $39^3$ in one batch. The central voxel of each patch is chosen via Tumor Sampling \cite{met_krivov2018tumor} strategy. We set the probability to choose voxel belonging to target mask to be $0.5$ for all experiments.

% We used 5-fold cross validation to split our training data patient-wise. After the train-predict process, we gather test predictions over the 5 splits to form the metric curve and compare experiment results. For the final report we used the whole training set to fit the models and reported the results on the holdout dataset.

\subsection{Contouring quality and time reduction}
\label{sec:clinical_exp}
Within a clinical experiment, we implemented the final model as a service which can process Dicom images and generate contours as Dicom RT files. This output was uploaded to a standard planning system and validated and adjusted (if needed) by experts there; we call these contours \emph{CNN-initialized}. In addition, the same cases were annotated manually in the same planning systems by the same four experts.

To perform the quality evaluation of our algorithm we introduced the following three types of comparisons. % (1 vs 3), the manual contour of one user comparing to the gold standard which was leave-one-out averaged contour of the three other users. The first setting allows us to measure the current inter-rater variability for a specific user. ($1^+$ vs 3), CNN-initialized contour of one user comparing to the same gold standard. In this setting we estimate the effect of CNN-initialization on the users. ($1^+$ vs $3^+$), the same as previous setting, but the gold standard was obtained averaging CNN-initialized contours for the three corresponding users. The last setting allows us to measure the level of additional standardization provided by CNN.

\begin{itemize}[label=$\bullet$]
    \item \textbf{1 vs 3} -- the manual contour of one user comparing to a ground truth estimation which is the averaged contour of the other users. This setting allows us to measure the current inter-rater variability for a specific user.
    \item $\mathbf{1^+}$ \textbf{vs 3} -- a CNN-initialized contour of one user comparing to the same ground truth as above. In this setting we estimate the effect of algorithm on the users.
    \item $\mathbf{1^+}$ \textbf{vs} $\mathbf{3^+}$ -- the same as previous setting, but the average contour was obtained using CNN-initialized contours for the three corresponding users. The last setting allows us to measure the level of additional standardization provided by CNN.
\end{itemize}

To investigate the differences in Dice scores we performed the Sign test for pairs of metrics (\textbf{1 vs 3}, $\mathbf{1^+}$ \textbf{vs 3}) and (\textbf{1 vs 3}, $\mathbf{1^+}$ \textbf{vs} $\mathbf{3^+}$), see Sec. \ref{sec:results}.

To evaluate a speed-up provided by our algorithm in routine clinical practice we compared times needed for two contouring techniques: manual delineation of the tumors and user adjustment of the CNN-initialized contours of the same tumors. The time spent on each task was recorded for all users and cases.

We didn't perform comparison types which include pure CNN generated contours, because AI could not be used in a treatment planing solely without user control and verification.

\section{Results}
\label{sec:results}

\subsection{Methods comparison on the hold-out dataset}

\begin{figure}[h!]

% \begin{minipage}[t]{0.485\textwidth}
    \includegraphics[width=\textwidth]{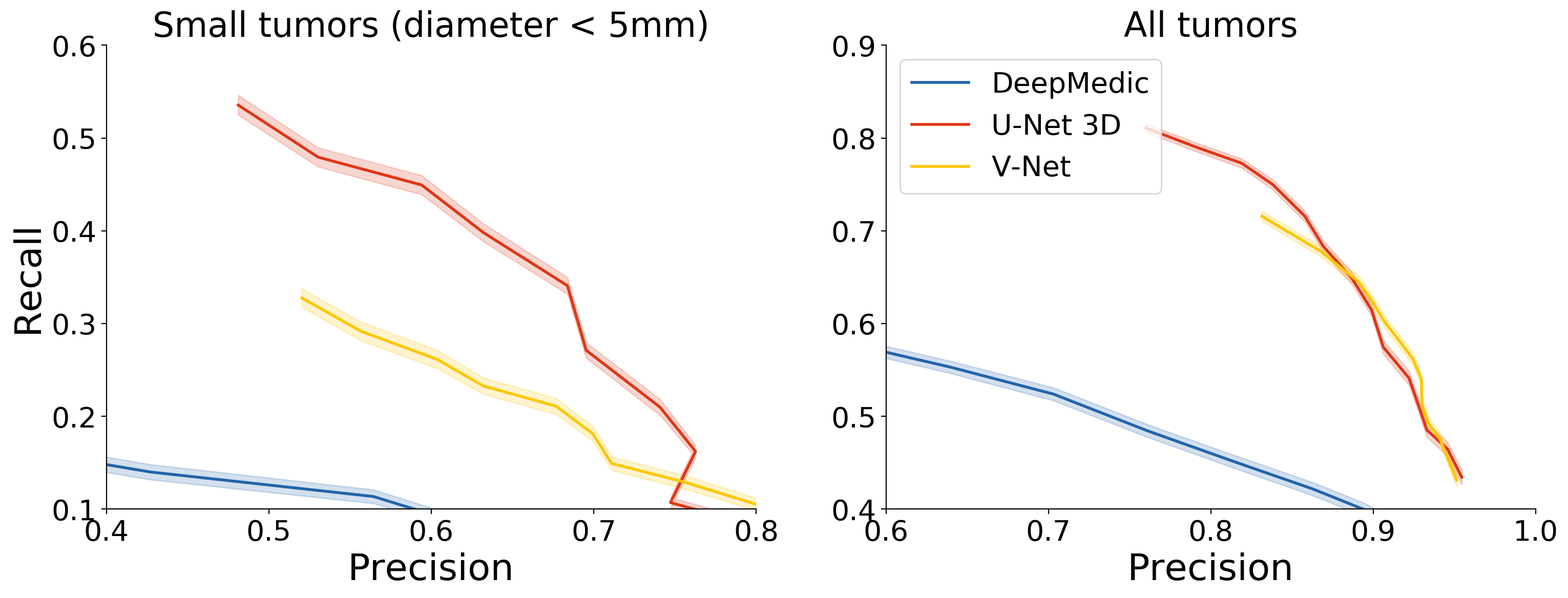}
    \caption{CNN models comparison. We zoomed all the PRC images from standard $\left[ 0; 1 \right]$ scale to better show some model or method had higher recall. We treated \emph{recall as a more important metric than precision} in our task: a radiologist spends few seconds on deleting miss-prediction but much more time on finding and delineating the tumor which CNN didn't predict.}\label{fig:models}
% \end{minipage}\hfill
% \begin{minipage}[t]{0.485\textwidth}

\end{figure}

\begin{figure}[h]

    \includegraphics[width=\textwidth]{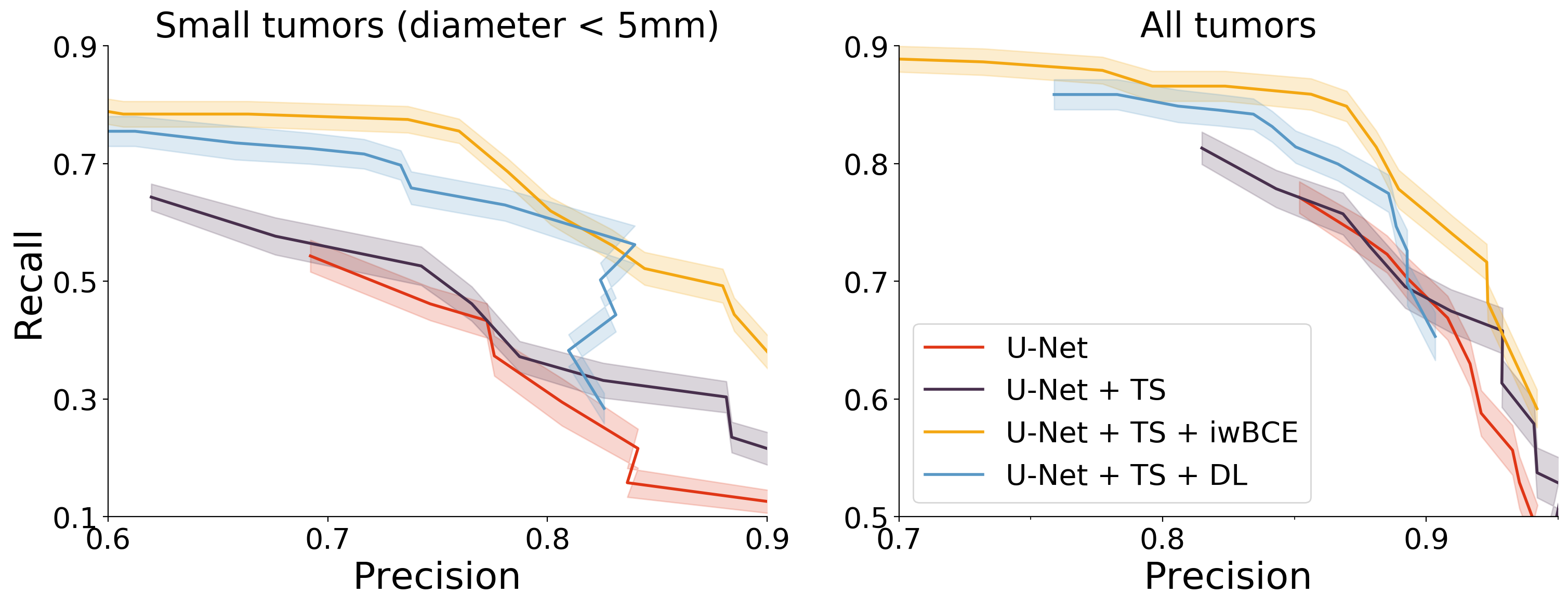}
    \caption{The best model with TS (Tumor Sampling) and then with iwBCE or DL (Dice Loss).}\label{fig:iwbce}
% \end{minipage}\hfill

\end{figure}

Firstly, we compared three network architectures, see Fig. \ref{fig:models}.
The results suggest the superiority of U-Net-like architectures over the DeepMedic in our task (see Fig. \ref{fig:models}). We made the architecture choice in favor of 3D U-Net and changed it in a minor way to fit our inference timings and memory requirements. We used this model for the subsequent experiments and the final model.

We also observed all the models perform poorly on the small tumors (Fig. \ref{fig:models}, left).
Within the second set of experiments, we aimed to improve recall for small lesions by adding Tumor Sampling and iwBCE to 3D U-Net, the best model from the first experiments. The proposed loss re-weighting strategy (see \ref{sec:iwbce}) reduced the number of missed small tumors by a factor of two with the same level of precision  (Fig. \ref{fig:iwbce}, left) and improve the network performance over all tumors (Fig. \ref{fig:iwbce}, right), achieving almost $0.9$ recall on the hold-out dataset. It slightly outperformed Dice loss function, so we used iwBCE to train our model for the clinical installation. 

The shaded area on the PRC plots shows 95\% confidence intervals of bootstrapped curves over 100 iterations choosing 80\% of the test patients every time. The median lesion-wise Dice score of 3D U-Net trained with Tumor Sampling and iwBCE is $0.84$ for the hold-out dataset.

\subsection{Clinical evaluation}

We observed better agreement between contours created by the expert and the reference one when the contours were initialized by CNN, even if the reference contour was generated completely manually. Tab. \ref{tab:abbr_sol} shows a reduction of inter-rater variability. Improvements for 3 out of 4 experts are statistically significant according to the Sign test p-values. The total median agreement increased from $0.924$ to $0.965$ in terms of Dice score. 

The automatic contours were generated and imported to the treatment planning system in less than one minute. The total median time needed to delineate a case manually was $10.09$ min., details for all four experts could be seen in tab. \ref{tab:time_reduct}. On average, the automatic algorithm speeds up the process of the delineation in $\mathbf{2.21}$ times with the median reduction of time of $5.53$ min. We observed speed-up for all users and for all cases they have delineated. We should note that acceleration plays more significant role in the cases of multiple lesions. The total median time needed to delineate a case with multiple metastases manually was $15.7$ min. (ranged from 15:20 to 44:00 in mm:ss). The automatic tumor segmentation speeded up the delineation of multiple lesions in $2.64$ times with median time reduction of $10.23$ min.

\begin{table}[h!]
    \centering
    \caption{Quality evaluation in tumor contouring. \emph{Case I} evaluated hypothesis that median difference between settings (\textbf{1 vs 3}) and ($\mathbf{1^+}$ \textbf{vs 3}) is equal to zero. \emph{Case II} evaluated the same hypothesis for settings (\textbf{1 vs 3}) and ($\mathbf{1^+}$ \textbf{vs} $\mathbf{3^+}$). 
    \emph{All data}  contains results for the consolidated set of experiments.}
    \begin{tabular}{l C{2cm} C{2cm} C{2cm}  C{2cm} C{2cm}}
        \toprule
        & \multicolumn{3}{c}{Median Dice Scores} & \multicolumn{2}{c}{p-values} \\
        \cmidrule(lr){2-4}
        \cmidrule(lr){5-6}
        & 1 vs 3 & $1^+$ vs 3 & $1^+$ vs $3^+$ & I & II \\
        \midrule
        User 1 & 0.938 & 0.947 & \textbf{0.969} & 2.85e-1 & 7.00e-6 \\
        User 2 & 0.930 & 0.941 & \textbf{0.968} & 7.01e-3 & 7.00e-6 \\
        User 3 & 0.915 & 0.920 & \textbf{0.934} & 2.29e-3 & 2.26e-3 \\
        User 4 & 0.918 & 0.935 & \textbf{0.968} & 1.40e-2 & 3.55e-2 \\
        \midrule
        All data  & 0.924 & 0.941 & \textbf{0.965} & 6.57e-4 & 3.61e-5 \\
        
        \bottomrule
    \end{tabular}
    \label{tab:abbr_sol}
\end{table}

\begin{table}[h!]
    \centering
    \caption{Time reduction in tumor delineation. Median time is given per one case.}
    \begin{tabular}{l c c c c}
        \toprule
        & \shortstack{Median \\ manual time\makebox[0pt][l]{$^{\ast}$}} & Range & \shortstack{Median \\ time reduction} & Range\\
        \midrule
        User 1 & 13:15 & 07:00 - 35:06 & 06:54 & 00:40 - 17:06 \\
        User 2 & 05:30 & 02:17 -- 15:20 & 02:16 & 00:48 -- 08:20 \\
        User 3 & 12:00 & 03:00 -- 44:00 & 09:00 & 01:00 -- 26:00 \\
        User 4 & 06:30 & 03:00 -- 23:30 & 05:27 & 03:00 -- 17:35 \\
        \midrule
        All data   & 10:05 & 02:17 -- 44:00 & 05:32 & 00:40 -- 26:00 \\
        \bottomrule
        \multicolumn{5}{l}{\footnotesize$^*$ the results are given in \textit{mm:ss}} \\
    \end{tabular}
    
    \label{tab:time_reduct}
\end{table}

\begin{figure}[h!]
    \centering
    \includegraphics[width=\linewidth]{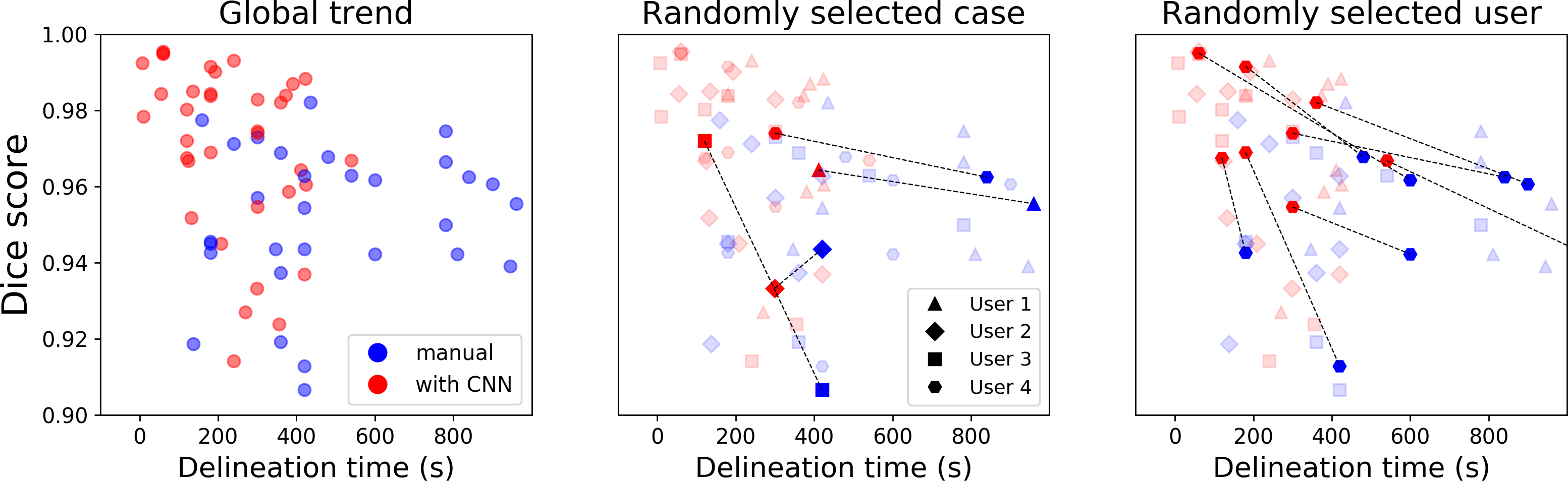}

    \caption{Plots of inter-rater agreement vs delineation time. \emph{Left}: each point corresponds to a pair lesion-user. Dice scores for blue dots (manual segmentation) were calculated using \textbf{1 vs 3} strategy, for red dots - \textbf{1 vs} $\mathbf{3^+}$. \emph{Central, right}: dashed lines connect two points for the same pair lesion-user for manual and CNN-initialized delineations. Note that we restricted both time-axis to the maximum of 1000 s and Dice-axis to the minimum of $0.9$, therefore few blue points were left outside the plot. }
    % Examples of applying different strategies of brain tumor delineation.
    \label{fig:time_dice}
\end{figure}

\begin{figure}[h!]
    \centering
    \includegraphics[width=\linewidth]{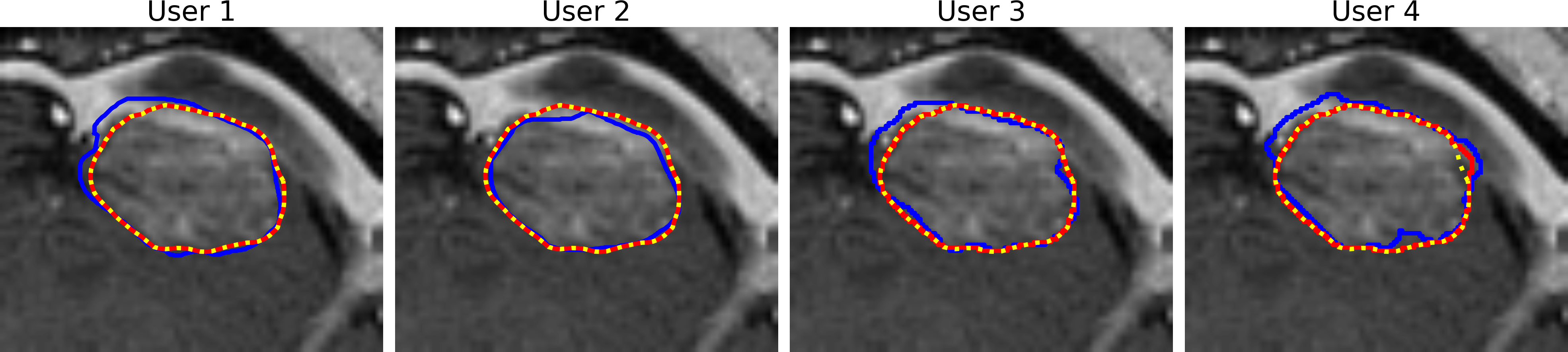}
    
    \includegraphics[width=\linewidth]{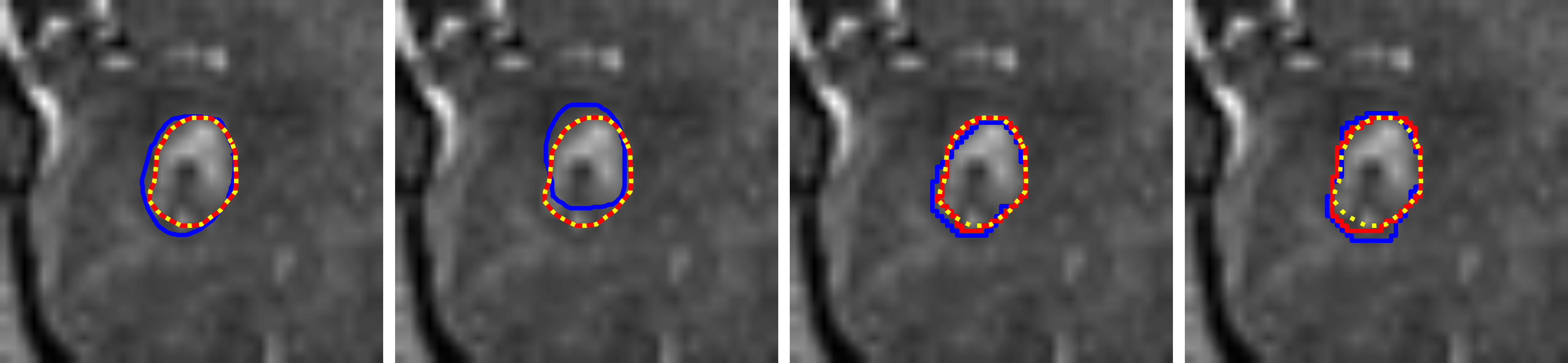}
    
    \includegraphics[width=\linewidth]{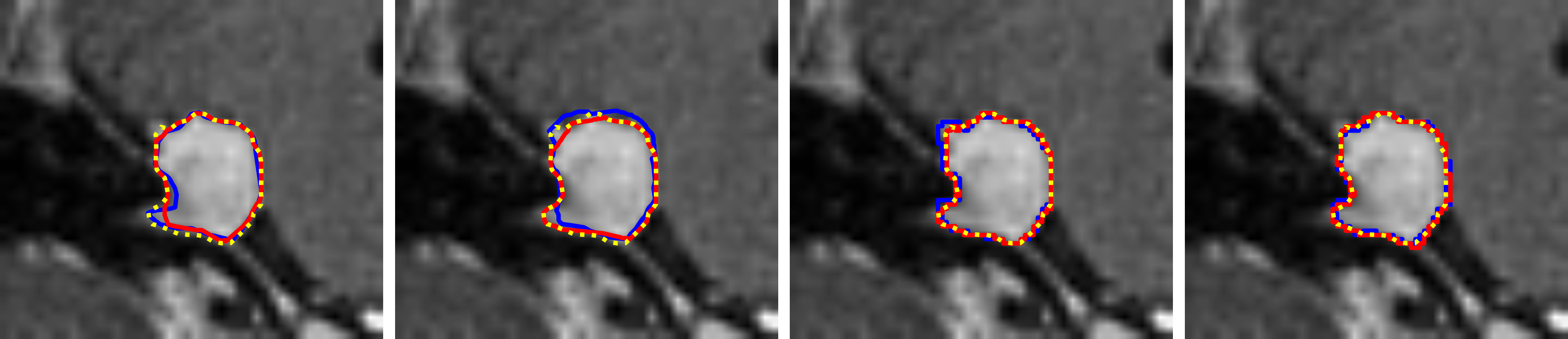}
    
    \includegraphics[width=\linewidth]{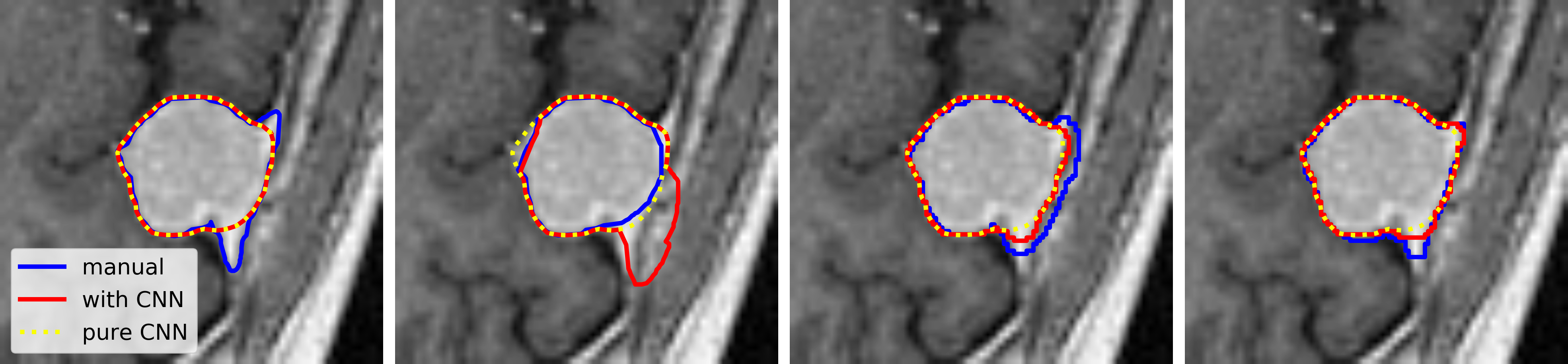}

    \caption{Segmentation results for two metastastic lesions, one schwannoma and one meningioma in vertical order.
    \textbf{Blue} corresponds to the manual contour,  \textbf{red} -- CNN-initialized contour with user's adjustment, \textbf{dashed yellow} --- pure CNN contour without user's adjustment.
    }
    \label{fig:contours}
\end{figure}

We also present quality-time plot (see fig. \ref{fig:time_dice}) for both manual and CNN-initialized techniques separately for each user and each case. One can distinguish the global trend of simultaneous improvement of inter-rater agreement and speedup of delineation time. Examples of different contouring techniques for all three types of lesions could be found on the fig. \ref{fig:contours}.

\section{Discussion}

For this study, we developed and successfully implemented a deep learning algorithm for automatic brain tumor segmentation into radiosurgery workflow. We demonstrated that our algorithm could achieve near expert-level performance, providing significant time savings in tumor contouring, and reducing the variability in targets delineation at the same time. We should note that within the clinical evaluation, the users initially delineated a case manually, and then they were asked to adjust the CNN-initialized contours of the same case. The adjustment of the CNN-initialized contours typically was performed in one day after manual delineation of the tumor. The fact that the experts had seen tumors previously might have a small impact on the results on the evaluation of time savings.    

We proposed a new loss function, called iwBCE, which has not been discussed in all the details. However, it seemed to be a promising approach to improve segmentation quality of modern deep learning tools. We aimed to continue research of the proposed method and compare it with state-of-the-art Dice loss in different setups and on different datasets.

\paragraph{Acknowledgements.} The Russian Science Foundation grant 17-11-01390 supported the development of the new loss function, computational experiments and article writing.

\bibliographystyle{splncs04}
\bibliography{main}

\begin{thebibliography}{10}
\providecommand{\url}[1]{\texttt{#1}}
\providecommand{\urlprefix}{URL }
\providecommand{\doi}[1]{https://doi.org/#1}

\bibitem{bratsmeta}
Bakas, S., Reyes, M., Jakab, A., Bauer, S., Rempfler, M., Crimi, A., Shinohara,
  R.T., Berger, C., Ha, S.M., Rozycki, M., et~al.: Identifying the best machine
  learning algorithms for brain tumor segmentation, progression assessment, and
  overall survival prediction in the brats challenge. arXiv preprint
  arXiv:1811.02629  (2018)

\bibitem{charron2018automatic}
Charron, O., Lallement, A., Jarnet, D., Noblet, V., Clavier, J.B., Meyer, P.:
  Automatic detection and segmentation of brain metastases on multimodal mr
  images with a deep convolutional neural network. Computers in Biology and
  Medicine  (2018)

\bibitem{unet3d}
{\c{C}}i{\c{c}}ek, {\"O}., Abdulkadir, A., Lienkamp, S.S., Brox, T.,
  Ronneberger, O.: 3d u-net: learning dense volumetric segmentation from sparse
  annotation. In: International conference on medical image computing and
  computer-assisted intervention. pp. 424--432. Springer (2016)

\bibitem{kamnitsas2017efficient}
Kamnitsas, K., Ledig, C., Newcombe, V.F., Simpson, J.P., Kane, A.D., Menon,
  D.K., Rueckert, D., Glocker, B.: Efficient multi-scale 3d cnn with fully
  connected crf for accurate brain lesion segmentation. Medical Image Analysis
  \textbf{36},  61--78 (2017)

\bibitem{krivov2018tumor}
Krivov, E., Kostjuchenko, V., Dalechina, A., Shirokikh, B., Makarchuk, G.,
  Denisenko, A., Golanov, A., Belyaev, M.: Tumor delineation for brain
  radiosurgery by a convnet and non-uniform patch generation. In: International
  Workshop on Patch-based Techniques in Medical Imaging. pp. 122--129. Springer
  (2018)

\bibitem{liu2017deep}
Liu, Y., Stojadinovic, S., Hrycushko, B., Wardak, Z., Lau, S., Lu, W., Yan, Y.,
  Jiang, S.B., Zhen, X., Timmerman, R., et~al.: A deep convolutional neural
  network-based automatic delineation strategy for multiple brain metastases
  stereotactic radiosurgery. Plos One  \textbf{12}(10),  e0185844 (2017)

\bibitem{liu2016automatic}
Liu, Y., Stojadinovic, S., Hrycushko, B., Wardak, Z., Lu, W., Yan, Y., Jiang,
  S.B., Timmerman, R., Abdulrahman, R., Nedzi, L., et~al.: Automatic metastatic
  brain tumor segmentation for stereotactic radiosurgery applications. Physics
  in Medicine \& Biology  \textbf{61}(24), ~8440 (2016)

\bibitem{BRATS}
Menze, B.H., Jakab, A., Bauer, S., Kalpathy-Cramer, J., Farahani, K., Kirby,
  J., Burren, Y., Porz, N., Slotboom, J., Wiest, R., et~al.: The multimodal
  brain tumor image segmentation benchmark (brats). IEEE Transactions on
  Medical Imaging  \textbf{34}(10),  1993--2024 (2015)

\bibitem{milletari2016v}
Milletari, F., Navab, N., Ahmadi, S.A.: V-net: Fully convolutional neural
  networks for volumetric medical image segmentation. In: 2016 Fourth
  International Conference on 3D Vision (3DV). pp. 565--571. IEEE (2016)

\bibitem{roques2014patient}
Roques, T.: Patient selection and radiotherapy volume definition—can we
  improve the weakest links in the treatment chain? Clinical Oncology
  \textbf{26}(6),  353--355 (2014)

\bibitem{sharp2014vision}
Sharp, G., Fritscher, K.D., Pekar, V., Peroni, M., Shusharina, N.,
  Veeraraghavan, H., Yang, J.: Vision 20/20: perspectives on automated image
  segmentation for radiotherapy. Medical Physics  \textbf{41}(5) (2014)

\bibitem{torrens2014standardization}
Torrens, M., Chung, C., Chung, H.T., Hanssens, P., Jaffray, D., Kemeny, A.,
  Larson, D., Levivier, M., Lindquist, C., Lippitz, B., et~al.: Standardization
  of terminology in stereotactic radiosurgery: Report from the standardization
  committee of the international leksell gamma knife society: special topic.
  Journal of Neurosurgery  \textbf{121}(Suppl\_2),  2--15 (2014)

\bibitem{van2010comparing}
Van~Ginneken, B., Armato~III, S.G., de~Hoop, B., van Amelsvoort-van~de Vorst,
  S., Duindam, T., Niemeijer, M., Murphy, K., Schilham, A., Retico, A.,
  Fantacci, M.E., et~al.: Comparing and combining algorithms for computer-aided
  detection of pulmonary nodules in computed tomography scans: the anode09
  study. Medical Image Analysis  \textbf{14}(6),  707--722 (2010)

\end{thebibliography}

\end{document}